\documentclass[journal]{IEEEtran}
\usepackage{fst}
\usepackage[T1]{fontenc}
\usepackage[utf8]{inputenc}

\usetikzlibrary{external}
\newcommand{\GL}[1]{#1}

\newcommand{\FS}[1]{#1}

\newcommand{\revA}[1]{#1}
\newcommand{\revB}[1]{#1}
\newcommand{\revC}[1]{#1}

\begin{document}

\title{One and Two Bit Message Passing for SC-LDPC Codes with Higher-Order Modulation}

\author{Fabian Steiner,~\IEEEmembership{Student~Member,~IEEE,} Emna Ben Yacoub, Bal\'azs Matuz,~\IEEEmembership{Member,~IEEE},\\ Gianluigi Liva,~\IEEEmembership{Senior Member,~IEEE}, Alexandre Graell i Amat,~\IEEEmembership{Senior Member,~IEEE}
\thanks{Fabian Steiner and Emna Ben Yacoub are with the Institute for Communications Engineering, Technical University of Munich (Email: \texttt{fabian.steiner@tum.de}, \texttt{emna.ben-yacoub@tum.de}).\newline Bal\'azs Matuz and Gianluigi Liva are with the Institute of Communications and Navigation, German Aerospace Center (DLR), Germany (E-mail: \texttt{balazs.matuz@dlr.de}, \texttt{gianluigi.liva@dlr.de}).\newline Alexandre Graell i Amat is with the Department of Electrical Engineering, Chalmers University of Technology, Gothenburg SE-41296, Sweden (E-mail: \texttt{alexandre.graell@chalmers.se}).\newline The work of A. Graell i Amat was supported by the Swedish Research Council (grant 2016-04253).}
}

\markboth{}{}%


\maketitle

\begin{abstract}
Low complexity decoding algorithms are necessary to meet data rate requirements in excess of \SI{1}{Tbps}. In this paper, we study one and two bit message passing algorithms for belief propagation decoding of low-density parity-check (LDPC) codes and analyze them by density evolution. The variable nodes (VNs) exploit soft information from the channel output. To decrease the data flow, the messages exchanged between check nodes (CNs) and VNs are represented by one or two bits. The newly proposed quaternary message passing (QMP) algorithm is compared asymptotically and in finite length simulations to binary message passing (BMP) and ternary message passing (TMP) for spectrally efficient communication with higher-order modulation and probabilistic amplitude shaping (PAS). To showcase the potential for high throughput forward error correction, spatially coupled LDPC codes and a target spectral efficiency (SE) of \SI{3}{bits/QAM symbol} are considered. Gains of about \SI{0.7}{dB} and \SI{0.1}{dB} are observed compared to BMP and TMP, respectively. The gap to unquantized belief propagation (BP) decoding is reduced to about \SI{0.75}{dB}. \revC{For smaller code rates, the gain of QMP compared to TMP is more pronounced and amounts to \SI{0.24}{dB} in the considered example.}
\end{abstract}


\section{Introduction}
\label{sec:intro}

\IEEEPARstart{O}{ptical} coherent transceivers with data rates of \GL{\SI{400}{Gbps}} are about to be installed in the field~\cite{ofc2018_postshow} and research already considers \SI{1}{Tb\GL{ps}}. These \GL{data} rates require sophisticated optical components, improved 
digital signal processing algorithms, and \ac{FEC} solutions that can cope with the high speed. While \ac{SD} decoders
are superior in terms of the \ac{NCG},
\ac{HD} decoders are appealing when low power consumption and high throughputs are of paramount importance. \ac{HD}-\ac{FEC} for optical communications is usually based on product-like codes with \ac{RS} or \ac{BCH} component codes of high rate, which can be efficiently decoded via \ac{BDD} (e.g., based on the syndrome). Spatially coupled \ac{HD}-\ac{FEC} constructions, such as staircase codes~\cite{smith_staircase_2012} or braided codes~\cite{feltstrom_braided_2009} achieve additional gains.

Recently, hybrid approaches based on concatenating an inner \ac{SD}-\ac{FEC} and an outer \ac{HD}-\ac{FEC} have received attention~\cite{magarini_concatenated_2010,zhang_low-complexity_2017}. These ideas have found their way into standards: the optical internet working forum established the 400G~ZR standard, a specification to transmit \GL{at \SI{400}{Gbps}} over data center interconnect links up to \SI{100}{km}, and agreed on an \ac{FEC} solution consisting of an inner \GL{Hamming code  and an outer  staircase code, where the inner code decoder is \ac{SD} and the outer code decoder is \ac{HD}},  with a total of $14.8\%$ overhead and a \ac{NCG} of \SI{10.8}{dB}.

To exploit the soft-information from the channel, while still only exchanging binary messages during the iterations of \ac{BDD}, the authors of~\cite{sheikh_binary_2019} weight the \ac{HD} output of the component decoders and recombine it with the soft-information from the channel, after which another \ac{HD} is made. \revB{Similar approaches were also considered in \cite{liga_novel_2019,lei_improved_2019}, where soft information from the channel is used to exploit particularly reliable and unreliable bits to improve the miscorrection-detection capability of the \ac{BDD} decoder.}

In~\cite{lechner_analysis_2012}, the authors present a one-bit \ac{BMP} algorithm for \ac{LDPC} codes. In particular, the  \ac{VN}  processor combines the soft channel message with scaled binary messages from the \acp{CN}, followed by a \ac{HD} step. The idea of passing binary messages dates back to the seminal work of Gallager~\cite{gallagerPhD}, where he presented algorithms that are now called Gallager A and Gallager B.

The internal decoder data flow $F$, \GL{defined as the} number of bits that are processed in each \ac{BP} iteration, is given by
 \begin{equation}
  F = \frac{2\cdot k_\tc \cdot q \cdot \bar d_{\mttv}}{R_\tc} = 2\cdot n_\tc \cdot q \cdot \bar d_{\mttv}
 \end{equation}
where $k_\tc$ is the number of information bits, $q$ is the number of bits used to represent a message, $R_{\tc}$ is the \ac{FEC} code rate and $\bar d_{\mttv}$ is average \ac{VN} degree. Observe that $n_\tc = k_{\tc}/R_{\tc}$ is the number of coded bits. Thus, a \ac{BMP} decoder ($q = 1$) allows to reduce the data flow $F$ by a factor of $q$ \FS{compared to a decoder using $q$ bits} \GL{to represent messages}. \revA{It hereby alleviates the routing congestion problem from which many high throughput, parallel \ac{LDPC} decoder architectures suffer~\cite{schlafer_new_2013}.}

The work in~\cite{BenY1902:Protograph} extends the \ac{BMP} algorithm to ternary messages.
The third message is an erasure that denotes complete uncertainty about the respective bit value. The algorithm, dubbed \ac{TMP} decoding, closely resembles algorithm~E from \cite{richardson_capacity}, except that it exploits soft-information at the \acp{VN}. 

These works study decoding for the \ac{BSC} or the \ac{AWGN} channel with \ac{BPSK}. For coherent, spectrally efficient optical communication, higher-order modulation formats are required and a dedicated code design should take the modulation into account.  This is particularly true for \ac{PAS}~\cite{bocherer_bandwidth_2015,bocherer_probabilistic_2019}, a coded modulation technique that uses a non-uniform distribution for the constellation points to operate close to the Shannon limit and allows flexible rate adaptation.

\revC{The main contributions of this paper are as follows:
\begin{itemize}
    \item We introduce a \ac{QMP} decoding algorithm as an extension of \ac{BMP} and \ac{TMP}. \ac{QMP} improves the decoding performance 
    for the considered examples by up to \SI{0.24}{dB} compared to \ac{TMP} for the same quantization resolution. The gain is particularly pronounced for low \ac{FEC} code rates.
    \item \FS{We complement the \ac{QMP} decoding algorithm by deriving its \ac{DE} to compute iterative decoding thresholds and obtain the optimal weighting factors required in the decoding algorithm.}
    \item We adapt \ac{BMP}, \ac{TMP}, and \ac{QMP} for higher-order modulation and \ac{PAS} targeting high-throughput decoding with spectrally efficient signaling and performance close to the Shannon limit by generalizing the \ac{DE} formulations to take the different \ac{BMD} bit channels into account. Previous work only considered \ac{BPSK} modulation. We also derive a simplified surrogate channel approach for the initialization of \ac{DE} for \ac{BMD} with higher-order modulation.
    \item We use \ac{BMP}, \ac{TMP}, and \ac{QMP} for \ac{SC-LDPC} codes and analyze their decoding performance in the finite length regime. The results show that an asymptotic design of the weights via \ac{DE} also yields very good performance at finite blocklength.
\end{itemize}}

This paper is organized as follows. In Sec.~\ref{sec:prelimin}, we introduce the system model, explain the principles of \ac{PAS} and review protograph-based \ac{SC-LDPC} codes. The one bit (\ac{BMP}) and two bit (\ac{TMP}, \ac{QMP}) message passing decoding approaches are described in Sec.~\ref{sec:decoding_approaches}. We develop \ac{DE} for the three algorithms and higher-order modulation with \ac{PAS} in Sec.~\ref{sec:de_bmp_tmp}. In Sec.~\ref{sec:numerical_results}, we present finite length simulation results and compare them to the asymptotic \ac{DE} predictions. We conclude in Sec.~\ref{sec:conclusions}.

\section{Preliminaries}
\label{sec:prelimin}

\subsection{Notation} 
We refer to the set of natural numbers as $\setN$; if zero is included, the set is denoted as $\setN_0$. The set of real numbers is $\setR$, the set of positive real numbers is $\setR^{+}$ and the set of non-negative real numbers is $\setR^+_{0}$. Other sets are denoted by calligraphic letters such as $\cA$. In general, vectors are written in bold font and are assumed to be row vectors, i.e., $\vx = (x_1,x_2,\dotsc, x_n)$. We denote \acp{RV} with upper case letters, e.g., $X$, and their realizations with lower case letters, e.g., $x$.
A discrete \ac{RV} $X$ has distribution $P_X$ on a discrete set $\cX$, e.g., for $x\in\cX$, $P_X(x)=\Pr\{X=x\}$.
A continuous \ac{RV} $X$ takes real values and has density $p_X$, e.g., $\int_{-\infty}^x p_X(\tau)\dif{\tau}=\Pr\{X\leq x\}$. 
The expectation of a RV $X$ is denoted by $\E{X}$.
Furthermore, $\entr(X)$, $\entr(X|Y)$ and $\I(X;Y)$, denote the entropy of the \ac{RV} $X$, the conditional entropy of the RV $X$ given the RV $Y$ and the mutual information of $X$ and $Y$, respectively.

\subsection{System Model}
\label{sec:system_model}

Previous works have shown that a discrete time \ac{AWGN} channel can be considered as an accurate model for dispersion-uncompensated, long-haul optical links~\cite{poggiolini_gn-model_2014}, where the noise contribution is dominated by amplified spontaneous emission noise from the erbium doped fiber amplifier. A transmitted modulation symbol $X_i$ is subject to Gaussian noise $N_i$ with zero mean and variance $\sigma^2$, and the received samples are
\begin{equation}
    Y_i = X_i + N_i, \quad i = 1,\dotsc, n.\label{eq:awgn_model}
\end{equation}
For ease of notation we drop the subscript $i$ whenever possible and we describe the signaling for one real dimension only, i.e., the channel inputs are from an $M$-ary \ac{ASK} set $\cX = \{\pm 1, \pm 3, \ldots, \pm (M-1)\}$, where $M$ is even. The extension to a two-dimensional $M^2$-\ac{QAM} constellation is straightforward: it can be obtained by the Cartesian product of two real-valued \ac{ASK} constellations. A four-dimensional dual polarized \ac{QAM} (DP-QAM) constellation can be obtained as the Cartesian product of two \ac{QAM} constellations and equivalently, as the Cartesian product of four \ac{ASK} constellations. 
The \ac{SNR} is defined as $\SNR = \E{X^2}/\sigma^2$.

\revB{To use binary \ac{FEC} codes with higher order modulation formats, we introduce a binary interface for the transmitted constellation points. We define the binary labeling $\chi\colon \cX \to \{0,1\}^m$ that assigns an $m$-bit binary label $\vb$ to each constellation point  $x$, i.e., $\vb = \chi(x)$. For \ac{BMD} the decoder uses this binary label to calculate a metric for each bit without taking their stochastic dependence into account.}

\subsection{Signaling and Achievable Rates}
\label{sec:achievable_rates}

The maximum rate at which reliable transmission over an \ac{AWGN} channel is possible is the Shannon capacity~\cite{shannon_mathematical_1948}.
It is achieved when the channel inputs are Gaussian distributed.
A uniform distribution on the constellation symbols in $\cX$ entails a loss in power efficiency compared to Gaussian signaling. The use of non-uniform probabilities is called \ac{PS}. 
Traditionally, the combination of \ac{PS} and \ac{FEC} was considered difficult (e.g., see~\cite[Sec.~6.2]{gallager1968}, \cite{forney_trellis_1992}).
Recently, the invention of \ac{PAS}~\cite{bocherer_bandwidth_2015} allowed an easy integration of \ac{PS} with \ac{FEC}. \ac{PAS} concatenates a shaping outer code called a \ac{DM}~\cite{schulte_constant_2016} and an \ac{FEC} inner code. This ``reverse concatenation'' was originally proposed for constrained coding in \cite{bliss1981circuitry,mansuripur1991enumerative}.
The \ac{PAS} architecture has three properties that distinguishes it from other \ac{PS} schemes. First, it integrates shaping with existing \ac{FEC}, second, it achieves the Shannon limit~\cite[Sec.~10.3]{bocherer2018principles},~\cite{amjad2018information}, and third, it \GL{allows} rate \GL{adaptation} by changing the probability distribution, while leaving the \ac{FEC} unchanged.

The \ac{DM}~\cite{schulte_constant_2016} realizes the non-uniform distribution. It takes $k_\tdm$ uniformly distributed
input bits and maps them to a length-$n$ sequence of symbols with the empirical distribution $P_A$. For PAS, the \ac{DM} output set is $\cA$ and the \ac{DM} rate is 
\begin{equation}
R_\tdm = k_\tdm/n\label{eq:Rdm}.
\end{equation}
The transmission rate $R_{\ttx}$ of \ac{PAS} is~\cite{bocherer_bandwidth_2015}
\begin{equation}
 R_{\ttx} = R_\tdm + 1 - (1-R_\tc)\cdot m \quad \text{[\si{\bpcu} (bits/channel use)]} \label{eq:Rtx_PAS}
\end{equation}
where $R_\tc$ is the code rate of the \ac{FEC} code.

An achievable rate for \ac{BMD} with \ac{SD} is given by~\cite{bocherer_probabilistic_2019}
\begin{equation}
 R_{\tbmd}(\SNR) = \left[\entr(\vB) - \sum_{k=1}^m \entr(B_k|Y)\right]^+\quad [\si{\bpcu}]\label{eq:rbmd}
\end{equation}
where $[\cdot]^+ = \max(0,\cdot)$. We introduce $R_{\tbmd}^{-1}(\cdot)$ as the inverse function. Its relation to other \ac{SD} \ac{FEC} metrics, e.g., the \ac{NGMI}~\cite{cho_probabilistic_2019} is explained in \cite{bocherer_probabilistic_2019}.

\ac{PAS} uses the \ac{BMD} soft-information 
\begin{align}
l_k(y) \triangleq \log\frac{P_{B_k|Y}(0|y)}{P_{B_k|Y}(1|y)}, \quad k = 1,\dotsc,m\label{eq:lch}
\end{align}
as decoder input, where $P_{B_k|Y}(b|y)$ is the probability that the $k$-th bit level of a constellation symbol is equal to $b$ for a  given $y$. We have
\begin{equation}
    P_{B_k|Y}(b|y) \propto \sum_{x\in\cX_k^b} p_{Y|X}(y|x) P_X(x)
\end{equation}
and $\cX_k^b \triangleq \{x\in\cX: [\chi(x)]_k = b\}$. The notation $[\vb]_k$ refers to the $k$-th component of the vector $\vb$.

\subsection{Protograph-Based Low-Density Parity-Check Codes}
\label{sec:protographs}

Binary \ac{LDPC} codes are binary linear block codes defined by an $ m_{\tc} \times n_{\tc} $ sparse parity-check matrix $\vH$. The code dimension is $k_{\tc} \geq n_{\tc} - m_{\tc}$. The Tanner graph of an \ac{LDPC} code is a bipartite graph $G = (\cV \cup \cC, \cE)$ consisting of $n_\tc$ \acp{VN} and $m_\tc$ \acp{CN}. The set $\cE$ of edges contains the element
$\mtte_{ij}$, where $\mtte_{ij}$ is an edge between \ac{VN} $\mttv_j\in\cV$ and \ac{CN} $\mttc_i\in\cC$. Note that $\mtte_{ij}$ belongs to the set $\cE$ if and only if the parity-check matrix element $h_{ij}$ (entry in the $i$-th row and $j$-th column of $\vH$) is equal to $1$.
The sets $\cN(\mttv_j)$ and $\cN(\mttc_i)$ denote the neighbors of \ac{VN} $\mttv_j$ and \ac{CN} $\mttc_i$,
respectively. The degree of a \ac{VN} $\mttv_j$ is denoted by $d_{\mttv_j}$ and it is the cardinality of the set $\cN(\mttv_j)$. Similarly, the degree of a \ac{CN} $\mttc_i$ is denoted by $d_{\mttc_i}$ and it is the cardinality of the set $\cN(\mttc_i)$.

For practical purposes, it is beneficial to impose
structure on an \ac{LDPC} code ensemble. Examples of structured \ac{LDPC} code ensembles are \ac{MET}~\cite{richardson_MET} and protograph-based ensembles~\cite{thorpe_protograph,divsalar_capacity-approaching_2009}.
Protograph-based ensembles are defined via a (typically small) basematrix $\vB$ of dimension $m_{\tp} \times n_{\tp}$ and elements in $\setN_0$. 
A basematrix may also be represented as a bipartite graph, called a protograph.
However, since the elements of the basematrix are not strictly binary, parallel edges are allowed and their numbers correspond to the respective entries in the basematrix. The Tanner graph of an \ac{LDPC} code \GL{can be  obtained by lifting a protograph}: through copy-and-permute operations copies of the protograph are generated and their edges are permuted such that connectivity constraints imposed by the basematrix are maintained \cite{thorpe_protograph}. A protograph-based \ac{LDPC} code ensemble $ \cC_{n_{\tc}}^{\vB} $ is defined by the set of all length-$n_{\tc}$ \ac{LDPC} codes whose Tanner graph is obtained by lifting $\vB$ by a factor of $Q$ such that $n_\tc = Q \cdot n_\tp$.\footnote{In this work, we do not consider \ac{LDPC} codes with state (i.e., punctured) \acp{VN}.} \FS{To distinguish the \acp{VN} and \acp{CN} in the protograph from those in the lifted parity-check matrix, we introduce the protograph \ac{VN} set $\cV_\tp = \{\mttV_1, \mttV_2, \dotsc, \mttV_{n_\tp}\}$ and \ac{CN} set $\cC_{\tp} = \{\mttC_1, \mttC_2, \dotsc, \mttC_{m_\tp}\}$. Every protograph \ac{VN} (\ac{CN}) identifies a \ac{VN} (\ac{CN}) type. We use the wording ``a type $\mttV_i$ \ac{VN}'' to identify a \ac{VN} in the lifted Tanner graph of type $\mttV_i$. We also use the convention that a \ac{VN} $\mttv_j$ in the Tanner graph is of type $\mttV_i$  if $\lceil j/Q \rceil = i$, i.e., consecutive blocks of $Q$ \acp{VN} in the final parity-check matrix are associated to a given type. The same applies to \acp{CN}.}

\subsection{Protograph-Based Spatially Coupled LDPC Codes}
\label{sec:sc_protograph}

Consider an \ac{SC-LDPC} code with a right-unterminated parity-check matrix~\FS{\cite{schmalen_spatially_2015}}
\begin{equation}
\vH = \vect{\vH_0(0)  &              &            &      \\
            \vH_1(0)  & \vH_0(1)     &            &   \\
            \vdots    & \vH_1(1)     & \vH_0(2)   &   \\
            \vH_{\mu}(0) & \vdots    & \vH_1(2)   &    \\
                      & \vH_{\mu}(1) & \vdots     & \ddots \\
                      &              & \vH_\mu(2) & \ddots \\
                      &              &            & \ddots}.\label{eq:sc_H}
\end{equation}
\GL{In \eqref{eq:sc_H},} $\mu$ \GL{denotes} the syndrome former memory of the \ac{SC-LDPC} code. The index in brackets denotes the \emph{spatial position}. If the matrices $\vH_i(s), i \in \{0, \dotsc, \mu\}$, are the same for all spatial positions $s \in \{0, \dotsc, S-1\}$, the \ac{SC-LDPC} code is called time-invariant and the index $s$ can be dropped. The dimension of the matrices $\vH_i(s)$ is $m_{\tc}^{\tSC}\times n_{\tc}^\tSC$.

Because of the diagonal structure of $\vH$,  a \ac{CN} is connected to at most $(\mu+1) n_{\tc}^{\tSC}$ \acp{VN}. \GL{This allows using} a window decoding approach~\cite{iyengar_windowed_2013} that reduces latency,  increases throughput, and makes \GL{\ac{SC-LDPC} codes} particularly interesting for optical communications~\cite{schmalen_spatially_2015}. 

\ac{SC-LDPC} codes are known to exhibit a phenomenon known as \emph{threshold saturation}~\cite{kudekar_threshold_2011} that allows to approach the \GL{bit-wise \ac{MAP}} decoding threshold of the underlying block code with \GL{(unquantized)} \ac{BP} decoding.

\ac{SC-LDPC} codes can be constructed from protographs and have the structure 
\begin{equation}
\vB = \vect{\vB_0     &              &            &      \\
            \vB_1     & \vB_0        &            &   \\
            \vdots    & \vB_1        & \vB_0      &   \\
            \vB_\mu   & \vdots       & \vB_1      &    \\
                      & \vB_{\mu}    & \vdots     & \ddots \\
                      &              & \vB_\mu    & \ddots \\
                      &              &            & \ddots}.\label{eq:sc_B}
\end{equation}

The protograph in \eqref{eq:sc_B} is then lifted by a factor of $Q$ to obtain the final parity-check matrix $\vH$.

For practical operation, the \ac{SC-LDPC} code is commonly terminated after a number of $S$ spatial positions. Due to this termination, a rate loss occurs that vanishes for large $S$. \FS{The resulting code rate is
\begin{equation}
R_{\tc} = 1 - \frac{\mu+S}{S}\frac{m_\tp^\tSC}{n_\tp^\tSC} = 1-\left(1+\frac{\mu}{S}\right)\frac{m_\tp^\tSC}{n_\tp^\tSC}\label{eq:sc_coderate}
\end{equation}
where the basematrices $\vB_0,\dotsc, \vB_\mu$ have dimensions $m_\tp^\tSC\times n_\tp^\tSC$. The overall size of the matrix $\vB$ is $m_\tp \times n_\tp = (\mu+S)m_\tp^\tSC \times n_\tp^\tSC S$.}

\section{Decoding Algorithms for One And Two Bit Message Passing}
\label{sec:decoding_approaches}

\GL{In this section, we first review  the \ac{BMP} and \ac{TMP} decoding algorithms introduced in \cite{lechner_analysis_2012} and \cite[Sec.~\ref{sec:bmp_tmp}]{BenY1902:Protograph}. In Sec.~\ref{sec:qmp}, we then present a new decoding algorithm that takes full advantage of $2$-bit messages, which we dub \ac{QMP}.}

\GL{For the described algorithms, w}e denote by  $ m_{\mttc \to \mttv}^{( \ell) }  $  the message sent from \ac{CN} $ \mttc $ to its neighboring \ac{VN} $ \mttv $ at the $\ell$-th iteration. Similarly, $ m_{\mttv \to \mttc}^{(\ell) }  $ is the message sent from \ac{VN} $\mttv$ to \ac{CN} $\mttc$. The soft information \GL{at the input of the decoder} for the $j$-th coded bit is denoted by $l_{\tdec,j}$ and  calculated according to \eqref{eq:lch}. 

\subsection{Binary and Ternary Message Passing}
\label{sec:bmp_tmp}

For \ac{BMP}, the exchanged messages are binary, i.e., $ m_{\mttv \to \mttc}^{(\ell)},  m_{\mttc \to \mttv}^{(\ell)} \in \cM_{\text{BMP}} \triangleq \{-1,+1\}$. For \ac{TMP}, the exchanged messages are ternary and we have $ m_{\mttv \to \mttc}^{(\ell)},  m_{\mttc \to \mttv}^{(\ell)} \in \cM_{\text{TMP}} \triangleq \{-1,0,+1\}$. A message value of zero indicates complete uncertainty about the respective bit.

In every decoding iteration, each \ac{VN} and \ac{CN} computes extrinsic messages that are forwarded to the neighboring nodes. Specifically, the message from \ac{VN} $\mttv$ to \ac{CN} $\mttc$ is obtained by combining the channel soft-information $l_{\tdec}$ with a weighted version of all other incoming \ac{CN} messages. Finally, a quantization function $\Psi\colon \setR \to \cM$ is applied to turn the result into binary and ternary messages for \ac{BMP} and \ac{TMP}, respectively. The weighting factors $w_{ij}^{(\ell)}$ are real valued and depend on the current iteration number. They can be obtained from the \ac{DE} analysis as shown in Sec.~\ref{sec:de_bmp_tmp}. The quantization function is
\begin{equation}
\Psi(x) = \begin{cases}
    +1, & x > 0\\
    -1, & x \leq 0
\end{cases}\label{eq:bmp_f}
\end{equation}
for \ac{BMP} and
\begin{equation}
\Psi(x) = \begin{cases}
+1, & x > T \\
0, & -T \leq x\leq T \\
-1, & x < -T
\end{cases}\label{eq:tmp_f}
\end{equation}
for \ac{TMP}. The equality signs in \eqref{eq:bmp_f} and \eqref{eq:tmp_f} are chosen such that ties are broken. Note that the threshold parameter $T\in\setR^+_0$ in \eqref{eq:tmp_f} \revA{depends on the \ac{SNR} and needs to be chosen for each signaling mode and iteration individually to minimize the decoding threshold. However, numerical studies reveal that a single value that is kept constant over the iterations entails almost no loss in performance. Therefore, we resort to this setting in the following.}

For the \ac{CN} to \ac{VN} update, a \ac{CN} sends the product of incoming messages from the other neighboring \acp{VN}. In the last iteration $\ell_{\tmax}$, the a-posteriori estimate of each codeword bit is calculated by taking a hard decision on the combined soft-information  from all \ac{CN} neighbors and the channel. 
The algorithmic procedure for \ac{BMP} and \ac{TMP} decoding is summarized in Algorithm~\ref{alg:bmp}. The weighting factors $w_{ij}^{(\ell)}$ have been derived as part of the \ac{DE} for BMP and TMP in \cite{BenY1902:Protograph}.

\begin{algorithm}[t]
\begin{algorithmic}
\State Set $m_{\mttv_j\to \mttc_i}^{(0)} = \Psi(l_{\tdec,j}), \forall j = 1, \ldots, n_\tc, \forall \mttc_i \in \cN(\mttv_j)$.
\State $\ell = 0$
\While{$\ell \leq \ell_{\tmax}$}
\State \texttt{// CN update}
\For{$i = 1, \dotsc, m{_\tc}$}
    \For{$\mttv_j \in \cN(\mttc_i)$}
        \State $m_{\mttc_i\to \mttv_j}^{(\ell) }  = \prod\limits_{\mttv_{j'} \in \cN (\mttc_i) \setminus \{\mttv_j\}} m_{\mttv_{j'}\to \mttc_i}^{({\ell-1})} $
    \EndFor
\EndFor
\State \texttt{// VN update}
\For{$j = 1, \dotsc, n{_\tc}$} 
    \For{$\mttc_i \in \cN(\mttv_j)$}
        \State $m_{\mttv_j\to \mttc_i}^{(\ell) }  = \Psi\left( l_{\tdec,j} + \sum\limits_{\mttc_{i'} \in \cN(\mttv_j)\setminus\{\mttc_i\}} w_{i'j}^{(\ell)} m_{\mttc_{i'} \to \mttv_j}^{(\ell)}\right)$
    \EndFor
\EndFor
\State $\ell = \ell + 1$
\EndWhile
\State \texttt{// Final codeword bit estimate}
\For{$j = 1, \dotsc, n{_\tc}$}
        \State $\hat c_{j}  = \frac{1}{2}-\frac{1}{2} \sign\left( l_{\tdec,j} + \sum\limits_{\mttc_{i} \in \cN(\mttv_j)} w_{ij}^{(\ell_{\tmax})} m_{\mttc_{i} \to \mttv_j}^{(\ell_{\tmax})}\right)$
\EndFor
\end{algorithmic}
\caption{BMP and TMP decoding.}
\label{alg:bmp}
\end{algorithm}

\subsection{Quaternary Message Passing}
\label{sec:qmp}

The \ac{TMP} algorithm of Sec.~\ref{sec:bmp_tmp} requires two bits per exchanged message. We now introduce a QMP decoding algorithm that requires the same number of bits per exchanged message, but allows a more granular quantization of the associated reliability soft-information.

The key idea of QMP is to distinguish between low and high reliability messages. The VN to CN and CN to VN messages, $ m_{\mttv \to \mttc}^{(\ell)}$ and  $m_{\mttc \to \mttv}^{(\ell)}$, take values in the quaternary alphabet $\cM_{\text{QMP}}\triangleq\{-\levH,-\levL,+\levL,+\levH\}$ and $\levL$ and $\levH$ correspond to messages with low and high reliability, respectively.
The  quantization  function is
\begin{align} \label{eq:functionf}
\Psi(x) =
\begin{cases}
-\levH, & x  \leq -T\\
-\levL, &  -T < x < 0 \\
+\levL, & 0 \leq x < T \\
+\levH, & x \geq T.
\end{cases}
\end{align}

 The \ac{QMP} decoding algorithm is summarized in Algorithm~\ref{alg:qmp}.  At the \acp{CN}, a min-sum decoding rule is employed. At the \acp{VN}, the incoming messages are weighted and combined with the channel soft-information. In contrast to \ac{BMP} and \ac{TMP}, two sets of weighting factors are needed for \ac{QMP} depending on the magnitude of the received message. The weights $ w_{ij,\levL}^{(\ell)}$ are used for messages with low reliability (i.e., $m_{\mttc\to \mttv} \in \{-\levL,+\levL\})$, whereas $w_{ij,\levH}^{(\ell)}$ are used for messages with high reliability (i.e., $m_{\mttc\to\mttv} \in \{-\levH,+\levH\}$).
\begin{algorithm}[t]
\begin{algorithmic}
\State Set $m_{\mttv_j\to \mttc_i}^{(0)} = \Psi(l_{\tdec,j}), \forall j = 1, \ldots, n_\tc, \forall \mttc_i \in \cN(\mttv_j)$.
\State $\ell = 0$
\While{$\ell \leq \ell_{\tmax}$}
\State \texttt{// CN update}
\For{$i = 1, \dotsc, m{_\tc}$}
    \For{$\mttv_j \in \cN(\mttc_i)$}
        \State $\begin{aligned}[t] m_{\mttc_i \to \mttv_j}^{(\ell) } =& \min\limits_{\mttv_{j'} \in \cN(\mttc_i) \setminus \{\mttv_j\}}|m_{\mttv_{j'} \to \mttc_{i}}^{(\ell-1) }| \cdot \\ &\prod\limits_{\mttv_{j'} \in \cN(\mttc_i) \setminus \{\mttv_j\}} \sign \left(m_{\mttv_{j'} \to \mttc_i}^{(\ell-1) }\right)
        \end{aligned}$
    \EndFor
\EndFor
\State \texttt{// VN update}
\For{$j = 1, \dotsc, n{_\tc}$} 
    \For{$\mttc_i \in \cN(\mttv_j)$}
     \State $ l_{\tav}^{(\ell) }= \sum\limits_{\mttc_{i'} \in \cN\left( \mttv_j\right) \setminus \{\mttc_i\} }\sign(m^{(\ell)}_{\mttc_{i'} \to \mttv_j}) w^{(\ell)}_{i'j,|m^{(\ell)}_{\mttc_{i'} \to \mttv_j}|}$
         \State $m_{\mttv_j\to \mttc_i}^{(\ell) }  = \Psi\left( l_{\tdec,j} + l_{\tav}^{(\ell) } \right)$
 \EndFor
\EndFor
\State $\ell = \ell + 1$
\EndWhile
\State \texttt{// Final codeword bit estimate}
\For{$j = 1, \dotsc, n{_\tc}$}
 \State $ l_{\tin}= \sum\limits_{\mttc_{i} \in \cN\left( \mttv_j\right) }\sign(m^{(\ell_{\tmax})}_{\mttc_{i} \to \mttv_j}) w^{(\ell_{\tmax})}_{ij,|m^{(\ell)}_{\mttc_{i} \to \mttv_j}|}$
        \State $\hat c_{j}  = \frac{1}{2}-\frac{1}{2}\sign\left( l_{\tdec,j} + l_{\tin} \right)$
\EndFor
\end{algorithmic}
\caption{QMP decoding.}
\label{alg:qmp}

\end{algorithm}

\section{Density Evolution Analysis for QMP}
\label{sec:de_bmp_tmp}

In the following, we describe \ac{DE} for \ac{QMP} and protograph-based \ac{LDPC} code ensembles. The purpose of \ac{DE} is twofold. First, it provides asymptotic decoding thresholds of the considered ensembles. Second, it allows to compute the optimal weighting factors for the \ac{VN} update that are needed for the decoding algorithm (optimal in an asymptotic sense, i.e., for infinite block length and lifting factor).

\subsection{Symmetry}
\label{sec:symmetry}

\ac{DE} analysis for binary \ac{LDPC} codes~\cite{richardson_capacity} assumes that the channel message and the extrinsic decoder messages fulfill the symmetry constraint
\begin{equation}
    p_{L|B}(l|0) = p_{L|B}(-l|1)\label{eq:def_symmetry_channel}
\end{equation}
where the \ac{RV} $L$ denotes the soft-information calculated from the channel output. In this case, one can assume that the all-zero codeword is transmitted and track the  probability of decoding failure over the iterations. As pointed out in~\cite{hou_capacity-approaching_2003}, for \ac{BMD} with higher-order modulation, the bit channels $p_{L_k|B_k}$ are generally not symmetric\footnote{Symmetry depends on the chosen labeling function $\chi$ and the input distribution $P_X$.}, where the \ac{RV} $L_k$ is defined as (cf.~\eqref{eq:lch})
\begin{align}
L_k \triangleq \log\frac{P_{B_k|Y}(0|Y)}{P_{B_k|Y}(1|Y)}, \quad k = 1,\dotsc,m.\label{eq:rv_ldec}
\end{align}
However, we can use \emph{channel adapters}~\cite{hou_capacity-approaching_2003} to introduce symmetrized
counterparts. This can be accomplished by using a pseudo-random, binary scrambling sequence at both the transmitter and receiver, which
modifies \eqref{eq:rv_ldec} as
\begin{equation}
\tilde L_k = L_k \cdot (1-2B_k)\label{eq:scrambling}.
\end{equation}
The resulting bit-channels $p_{\tilde L_k|B_k}$ are symmetric, i.e., we have
\begin{equation}
    p_{\tilde L_k|B_k}(l|0) = p_{\tilde L_k|B_k}(-l|1).
\end{equation}

\subsection{Initialization of Density Evolution for Different Bit Channels}
\label{sec:initialization_de_tmp}

\FS{We associate with each \ac{VN} type a bit level. Let $\phi(j)$ be the bit level on which the \acp{VN} of type $\mttV_j$ are mapped and let $\cV_\tp^{(k)}\subseteq \cV_\tp$ be the subset of protograph \acp{VN} that are mapped to the $k$-th bit level.} 
We assume that the number $n_\tp$  of \acp{VN} in the protograph is an integer multiple of $m$, such that each bit level is assigned to $n_\tp/m$ \acp{VN}. 

Let $p_{\mttm}^{(\ell)}( i,j)$ be the probability that the message sent from $ \mttV_{j} $ to  $ \mttC_{i} $  at the $\ell$-th iteration on one of the $ b_{ij} $ edges connecting $ \mttV_{j} $ to $ \mttC_{i} $ is equal to $\mttm \in \{-\levH,-\levL,+\levL\}$.
To initialize \ac{DE}, we calculate the initial message probabilities as
\begin{align}
   p_{-\levH}^{(0)}(i,j)  &= \int_{-\infty}^{-T} p_{\tilde L_{\phi(j)}|B_{\phi(j)}}(l|0) \dif{l}\label{eq:qmp_int_-H}\\            
      p_{-\levL}^{(0)}(i,j) &= \int_{-T}^{0} p_{\tilde L_{\phi(j)}|B_{\phi(j)}}(l|0) \dif{l}\label{eq:qmp_int_-L}\\
     p_{+\levL}^{(0)}(i,j) &= \int_{0}^{T} p_{\tilde L_{\phi(j)}|B_{\phi(j)}}(l|0) \dif{l}\label{eq:qmp_int_+L}.
\end{align}

The integrals in \eqref{eq:qmp_int_-H}--\eqref{eq:qmp_int_+L} do not allow a closed form solution, but can be calculated by means of Monte Carlo simulations or transformations of \acp{RV}. Note that the above calculations need to be performed only once.

\begin{figure*}[ht]
\footnotesize
\centering
\subfloat[][Uniform]{\includegraphics{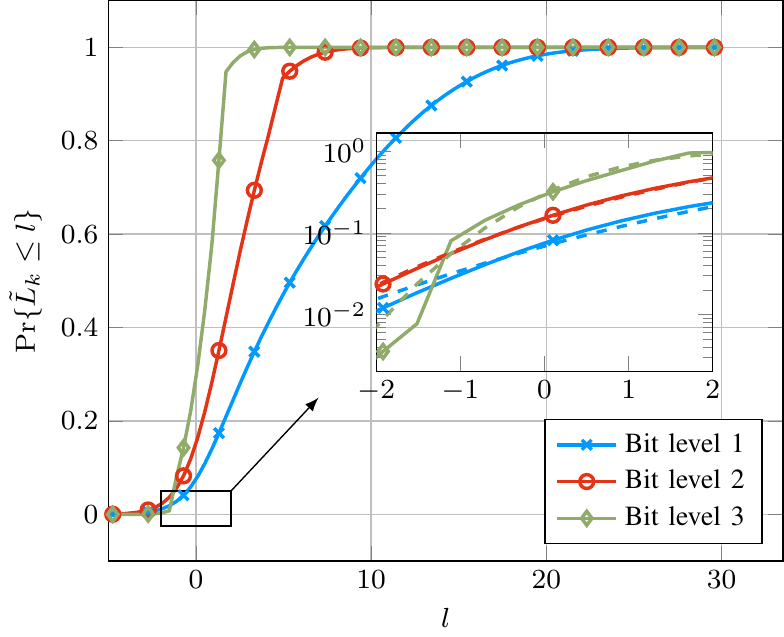}}
\subfloat[][PS]{\includegraphics{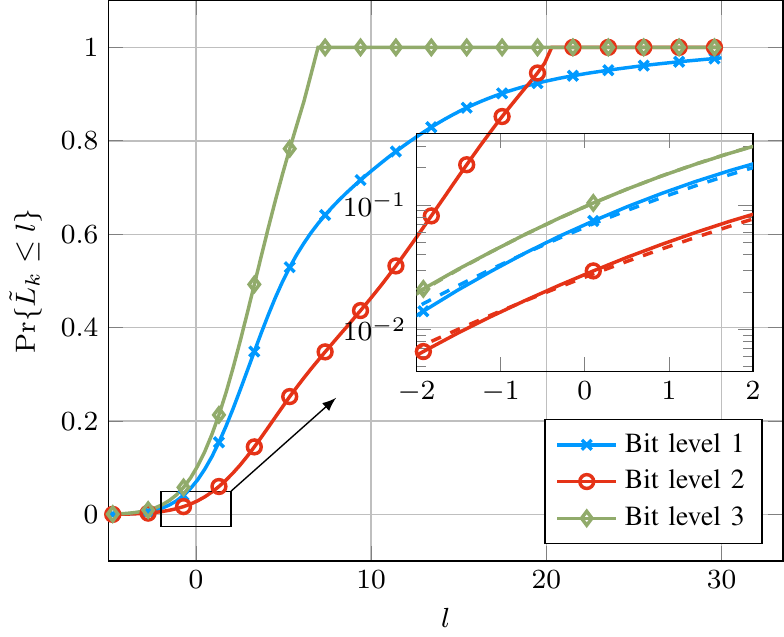}}
\caption{Comparison of \ac{CDF} plots for 8-ASK with uniform and \ac{PS} signaling. Both scenarios are for $\SNR = \SI{9}{dB}$. \ac{PS} signaling uses an \ac{MB} distribution with entropy $\entr(X) = \SI{2.25}{bits}$. The dashed lines in the insets denote the CDFs obtained via the surrogate approach.}
\label{fig:cdf_plots}
\end{figure*}

\revA{In Fig.~\ref{fig:cdf_plots}, we show the \acp{CDF} $\Pr\{\tilde L_k \leq l\}$ for 8-\ac{ASK} with uniform and PS signaling obtained via Monte Carlo simulations. The \acp{CDF} can be used to calculate \eqref{eq:qmp_int_-H}--\eqref{eq:qmp_int_+L} as
\begin{align}
   p_{-\levH}^{(0)}(i,j)  &= \Pr\{\tilde L_{\Phi(j)} \leq -T\},\\            
   p_{-\levL}^{(0)}(i,j) &= \Pr\{\tilde L_{\Phi(j)} \leq 0\} - \Pr\{\tilde L_{\Phi(j)} \leq -T\},\\
   p_{+\levL}^{(0)}(i,j) &= \Pr\{\tilde L_{\Phi(j)} \leq T\} - \Pr\{\tilde L_{\Phi(j)} \leq 0\}.
\end{align}
}

\subsection{Density Evolution for QMP}
\label{sec:algorithmic_description_de_qmp}

Similarly to before, let $ q_{\mathtt{m}}^{(\ell)}( i,j)$ denote the probability that the message sent from $\mttC_i$ to  $\mttV_j$  at the $\ell$-th iteration is equal to $\mathtt{m}\in\cM_{\tqmp}$.
\begin{enumerate}
	\item \textbf{Initialization.} For $ j= 1, 2, \ldots, n_{\tp} $ and $ i=1, 2, \ldots, m_{\tp}$ with $ b_{ij} \neq 0 $, compute $p^{(0)}_{-\levH}(i,j)$, $p^{(0)}_{-\levL}(i,j)$  and $p^{(0)}_{+\levL}(i,j)$ according to \eqref{eq:qmp_int_-H}, \eqref{eq:qmp_int_-L} and \eqref{eq:qmp_int_+L}.
 \item \textbf{For} $ \ell =1, 2, \ldots, \ell_{\tmax} $ repeat the following three update steps

\medskip
		\textbf{Check to variable update.} For $ j= 1, 2, \ldots, n_{\tp} $ and $ i=1, 2, \ldots, m_{\tp}$, if $ b_{ij} \neq 0 $ then compute
	\begin{equation}
	\footnotesize
	\begin{aligned}
	&q^{(\ell)}_{-\levH}(i,j)=\frac{1}{2} \left[ \prod\limits_{ b_{is} \neq 0}\left(1-	p^{(\ell -1 )}_{-\levL}(i,s)-	p^{(\ell -1)}_{+\levL}(i,s)\right)^{b_{is}- \delta_{sj}} \right. \\
	& \left. - \prod\limits_{ b_{is} \neq 0}\left(1-2p^{(\ell -1 )}_{-\levH}(i,s)-	p^{(\ell -1 )}_{-\levL}(i,s)-	p^{(\ell -1)}_{+\levL}(i,s)\right)^{b_{is}- \delta_{sj}} \right]
	\end{aligned}
\end{equation}

	\begin{equation}
	\footnotesize
	\begin{aligned}
	&q^{(\ell)}_{-\levL}(i,j)= \frac{1}{2} \left[1 - \prod\limits_{b_{is} \neq 0}\left(1-	p^{(\ell -1 )}_{-\levL}(i,s)-	p^{(\ell -1)}_{+\levL}(i,s)\right)^{b_{is}- \delta_{sj}}\right. \\ 
	& - \prod\limits_{b_{is} \neq 0}\left(1-2p^{(\ell -1 )}_{-\levH}(i,s)-	2p^{(\ell -1 )}_{-\levL}(i,s)\right)^{b_{is}- \delta_{sj}} \\
	& \left. + \prod\limits_{b_{is} \neq 0}\left(1-2p^{(\ell -1 )}_{-\levH}(i,s)-	p^{(\ell -1 )}_{-\levL}(i,s)-	p^{(\ell -1)}_{+\levL}(i,s)\right)^{b_{is}- \delta_{sj}} \right]
		\end{aligned}
\end{equation}
	\begin{equation}
	\footnotesize
	\begin{aligned}
	&q^{(\ell)}_{+\levL}(i,j)=\frac{1}{2}\left[1 - \prod\limits_{b_{is} \neq 0}\left(1-	p^{(\ell -1 )}_{-\levL}(i,s)-	p^{(\ell -1)}_{+\levL}(i,s)\right)^{b_{is}- \delta_{sj}}  \right. \\ 
	& +\prod\limits_{ b_{is} \neq 0}\left(1-2p^{(\ell -1 )}_{-\levH}(i,s)-	2p^{(\ell -1 )}_{-\levL}(i,s)\right)^{b_{is}- \delta_{sj}} \\
	& \left. - \prod\limits_{b_{is} \neq 0}\left(1-2p^{(\ell -1 )}_{-\levH}(i,s)-	p^{(\ell -1 )}_{-\levL}(i,s)-	p^{(\ell -1)}_{+\levL}(i,s)\right)^{b_{is}- \delta_{sj}} \right].
	\end{aligned}
\end{equation}
\medskip

		\textbf{Variable to check update.} For $ j= 1, 2, \ldots, n_{\tp} $ and $ i=1, 2, \ldots, m_{\tp}$, if $ b_{ij} \neq 0 $, compute
	\begin{align} 
	\begin{split}
	p^{(\ell)}_{-\levH}(i,j)=& \sum\limits_{z} \Pr\left\{  L^{(\ell)}_{\tav} = z \right\} \Pr\{\tilde L_{\phi(j)} \leq -T-z\}  \label{eq:pldpc_qmp_de_v2c_-H}
	\end{split}\\
	\begin{split}
	p^{(\ell)}_{-\levL}(i,j)=& \sum\limits_{z} \Pr\left\{  L^{(\ell)}_{\tav} = z \right\}\Pr\{-T-z<\tilde L_{\phi(j)} < -z\}  \label{eq:pldpc_qmp_de_v2c_-L}
	\end{split} \\
	\begin{split}
	p^{(\ell)}_{+\levL}(i,j)=& \sum\limits_{z} \Pr\left\{  L^{(\ell)}_{\tav} = z \right\} \Pr\{-z \leq \tilde L_{\phi(j)} < T-z\}  \label{eq:pldpc_qmp_de_v2c_+L}
	\end{split}
	\end{align}
		
		where $ L^{(\ell)}_{\tav} $ is a \ac{RV} representing the sum of the \acp{LLR} of the $d_{\mttv_j} - 1 $ \ac{CN} messages at the input of $\mttV_{j} $ at the $ \ell$-th iteration. We have
			\begin{equation}
		\begin{aligned}
		&\Pr\left\{  L^{(\ell)}_{\tav}= z \right\} = \sum\limits_{\vu, \vv,\vt} \prod\limits_{b_{sj}\neq 0} 
			{\textstyle  \binom{b_{sj} - \delta_{si}}{u_{s},v_{s},t_{s},b_{sj} - \delta_{si} - u_{s} - v_{s}-t_{s}}} \cdot \\
		& q^{(\ell)}_{-\levH}(s,j)^{u_{s}} q^{(\ell)}_{-\levL}(s,j)^{v_{s}}  q^{(\ell)}_{+\levL}(s,j)^{t_{s}} \cdot \\
		& \left(1-q^{(\ell)}_{-\levH}(s,j) -q^{(\ell)}_{-\levL}(s,j)-q^{(\ell)}_{+\levL}(s,j)\right)^{b_{sj} - \delta_{si} - u_{s} - v_{s}-t_{s}}
		\end{aligned}
		\end{equation}
		where the outer sum is over all integer vector triplets $\vu, \vv$ and  $\vw$ for which 
		\begin{equation}
		    \begin{aligned}
		\sum\limits_{e=1}^{m_{\tp}} &\left[w_{ej,\levL}^{(\ell)}(t_{e}-v_{e}) + \right. \\
	&\left. w_{ej,\levH}^{(\ell)}(b_{ej}-\delta_{ei}-2u_{e}-v_{e}-t_{e})\right] = z
		\end{aligned}
				\end{equation}
		with
		\begin{equation}\label{eq:DLej}
		w_{ej,\levL}^{(\ell)} = \ln\left( \dfrac{q^{(\ell)}_{+\levL}(e,j)}{	q^{(\ell)}_{-\levL}(e,j)}\right)
		\end{equation}
		and 
		\begin{equation}\label{eq:DHej}
		w_{ej,\levH}^{(\ell)} = \ln\left( \dfrac{1-q^{(\ell)}_{-\levH}(e,j) -q^{(\ell)}_{-\levL}(e,j)  -	q^{(\ell)}_{+\levL}(e,j)}{	q^{(\ell)}_{-\levH}(e,j)}\right).
		\end{equation}
		Specifically, the entries $u_{s} $, $v_{s}$, and $t_{s} $ represent the number of messages $-\levH$, $-\levL$ and $+\levL$, respectively, that $\mttC_{s} $ sends to $ \mttV_{j} $ on $ b_{sj} -\delta_{si}  $ of the $ b_{sj} $ edges connecting $ \mttC_{s} $ to $ \mttV_{j}$. Thus, for $s = 1, 2, \ldots, m_{\tp}$ we have
		$0 \leq u_{s}, v_{s}, t_{s} $ and $u_{s}+v_{s}+t_{s} \leq b_{sj}-\delta_{si}$.
	\medskip
	
		\textbf{A-posteriori update.} For $ j= 1, 2, \ldots, n_{\tp} $, compute
	
		\begin{equation}
	 P_{\tapp}^{(\ell)}(j)= 
		\sum\limits_{z} \Pr\left\{  L^{(\ell)}_{\tin} = z \right\}  \Pr\left\{\tilde L_{\phi(j)} \leq -z\right\}
		\end{equation}
		where $ {L}^{(\ell)}_{\tin} $ is a \ac{RV} representing the sum of the \acp{LLR} of all incoming $d_{\mttv_j}$ \ac{CN} messages at the input of $\mttV_{j} $ at the $ \ell$-th iteration.	We have	
			\begin{equation}
		\begin{aligned}
			&\Pr\left\{  L^{(\ell)}_{\tin}= z \right\} = \sum\limits_{{\bm{u}}, {\bm{v}}, {\bm{t}}} \prod\limits_{b_{sj}\neq 0} 
		{\textstyle \binom{b_{sj} }{{u}_{s},{v}_{s},{t}_{s},b_{sj} - {u}_{s} - {v}_{s}-{t}_{s}}} \cdot \\
		& q^{(\ell)}_{-\levH}(s,j)^{{u}_{s}} q^{(\ell)}_{-\levL}(s,j)^{{v}_{s}}  q^{(\ell)}_{+\levL}(s,j)^{{t}_{s}} \cdot \\
		& \left(1-q^{(\ell)}_{-\levH}(s,j) -q^{(\ell)}_{-\levL}(s,j)-q^{(\ell)}_{+\levL}(s,j)\right)^{b_{sj}  - {u}_{s} - {v}_{s}-{t}_{s}}
		\end{aligned}
		\end{equation}
			where the outer sum is over all integer vector triplets ${\bm{u}},{\bm{v}}$ and  ${\bm{t}}$ for which 
				\begin{equation}
		\begin{aligned}
	\sum\limits_{e=1}^{m_\tp} &\left[w_{ej,\levL}^{(\ell)}({t}_{e}-{v}_{e})+
	 w_{ej,\levH}^{(\ell)}(b_{ej}-2{u}_{e}-{v}_{e}-{t}_{e})\right] = z
	\end{aligned}
				\end{equation}
		where $	w_{ej,\levL}^{(\ell)}$ and $w_{ej,\levH}^{(\ell)}$ are given in \eqref{eq:DLej} and \eqref{eq:DHej}. The vector elements $  {u}_{s} $, $ {v}_{s} $, and ${t}_{s} $ represent the number of messages $ -\levH$, $-\levL$ and $+\levL$, respectively, that $ \mttC_{s} $ sends to $ \mttV_{j} $ on  the $ b_{sj} $ edges connecting $ \mttC_{s} $ to $ \mttV_{j}$. Thus, for $s = 1, 2, \ldots, m_\tp$ we have
		$0 \leq {u}_{s}, {v}_{s}, {t}_{s} $ and ${u}_{s}+{v}_{s}+{t}_{s} \leq b_{sj}$.

	\end{enumerate}

\begin{myremark}
With a slight abuse of notation, the weighting factors for \ac{QMP} in  Algorithm~\ref{alg:qmp} are calculated from those derived in this section as
\begin{align}
 w_{ij,\levH}^{(\ell)} \Leftrightarrow w_{\lceil i/Q\rceil \lceil j/Q\rceil,\levH}^{(\ell)}\\
 w_{ij,\levL}^{(\ell)} \Leftrightarrow w_{\lceil i/Q\rceil \lceil j/Q\rceil,\levL}^{(\ell)}
\end{align}
for $i= 1, \dotsc, m_{\tc}$ and $j = 1, \dotsc, n_{\tc}$.
\end{myremark}
\revA{
\begin{myremark}
Practical decoder implementations may also quantize the weighting factors or share the same weighting factors for several iterations. Similar decoding approaches for product codes~\cite{sheikh_binary_2019} suggest only minimal performance degradation in this case~\cite{fougstedt_energy-efficient_2019}.
\end{myremark}
}

\subsection{Surrogate Channel Approach for the Density Evolution Initialization} 

As an alternative to Monte Carlo simulations, we may resort to a \emph{surrogate channel} approach~\cite{peng2006,franceschini2006,sason_universal_2009} to approximate the required input probabilities~\eqref{eq:qmp_int_-H}--\eqref{eq:qmp_int_+L}. For this, the bit-channels $p_{\tilde L_k|B_k}$ are replaced by ``equivalent'' \ac{AWGN} channels with uniform binary inputs for which the derivation of the \acp{CDF} is easier. We establish their ``equivalence''\footnote{The term ``equivalence'' is not meant to have a strict information theoretic meaning in this context. Rather, this term refers to the observation that both types of threshold evaluations yield similar results numerically.} by requiring that the  channel and its surrogate have the same channel uncertainty. Let the surrogate be $\breve Y_k = \breve X_k + \breve N_k$ with $\breve X_k \in \{-1,+1\}$ and $\breve N_k \sim \cN(0,\breve \sigma_k^2)$ for $k = 1, \ldots, m$. For each \ac{SNR}, we calculate the set of equivalent channel parameters
\begin{align}
\breve\sigma_k^2: \entr(\breve B_k|\breve Y) = \entr(B_k|Y), \quad k = 1, \ldots, m.
\end{align}
For \ac{QMP} we obtain the expressions
\begin{align} 
\begin{split}
	p^{(0)}_{-\levH}(i,j)=Q\left( \dfrac{T+\mu_{\tch,\phi(j)}}{\sigma_{\tch,\phi(j)}}\right)\label{eq:qmp_int_-H_surrogate}
\end{split}\\
\begin{split}
	p^{(0)}_{-\levL}(i,j)=	Q\left( \dfrac{\mu_{\tch,\phi(j)}}{\sigma_{\tch,\phi(j)}}\right)  -  Q\left( \dfrac{T+\mu_{\tch,\phi(j)}}{\sigma_{\tch,\phi(j)}}\right)\label{eq:qmp_int_-L_surrogate}
\end{split}\\ 
\begin{split}
	p^{(0)}_{+\levL}(i,j)=	Q\left( \dfrac{-T+\mu_{\tch,\phi(j)}}{\sigma_{\tch,\phi(j)}}\right)  -  Q\left( \dfrac{\mu_{\tch,\phi(j)}}{\sigma_{\tch,\phi(j)}}\right)\label{eq:qmp_int_+L_surrogate}
\end{split}
\end{align}
where $\mu_{\tch,k} = 2/{\breve \sigma_k^2}$, $\sigma_{\tch,k}^2 = 4/{\breve \sigma_k^2}$, and $Q(\cdot)$ is the standard normal Gaussian tail probability, i.e., 
\begin{align}
Q(x) = \int_x^\infty (1/\sqrt{2\pi}) \exp(-\tau^2/2) \dif{\tau}.    
\end{align}
\revA{In Fig.~\ref{fig:cdf_plots}, we show the approximations of the true \acp{CDF} by the surrogate approach (dashed lines). A close match of the true \acp{CDF} and their approximations is observed.}

\subsection{Density Evolution of Window Decoding for Spatially Coupled Low-Density Parity-Check Codes}
\label{sec:tmp_den_evo_window}

We follow the approach of \cite{iyengar_windowed_2012} to determine the decoding threshold of protograph-based \ac{SC-LDPC} code ensembles for window decoding. For this, we apply the \ac{DE} analysis of Sec.~\ref{sec:de_bmp_tmp} for the respective decoding algorithm on a protograph matrix $\vB_{[1:W, 1:W]}$ that has been derived from \eqref{eq:sc_B} for a given decoding window size of $W$ with $\mu +1 \leq W \leq L$. The notation $\vB_{[1:W, 1:W]}$ denotes the block matrix of size $W\times W$ that is formed from the first $W$ block rows and $W$ block columns of $\vB$. For instance, for $\mu = 2$ and $W = 4$ we have
\begin{equation}
\vB_{[1:4,1:4]} = \vect{
\vB_0 & \zeros        &  \zeros       &   \zeros    \\
\vB_1 & \vB_0   &   \zeros      &   \zeros   \\
\vB_2 & \vB_1   & \vB_0   &   \zeros    \\
\zeros & \vB_2   & \vB_1   & \vB_0}.
\end{equation}

Convergence of the window decoder is declared when the probability of decoding error for the \acp{VN} in the first block column is (approximately) zero. The respective decoding threshold is referred to $\SNR_{\tth}^{\tbmp}$ for \ac{BMP}, $\SNR_{\tth}^{\ttmp}$ for \ac{TMP}, and $\SNR_{\tth}^{\tqmp}$ for \ac{QMP}, respectively.

\section{Numerical Results}
\label{sec:numerical_results}

We investigate the following signaling modes. \revC{The first one operates at \SI{1.0}{bpcu}, whereas the others operate at a \ac{SE} of \SI{1.5}{bpcu}:}
\begin{enumerate}
\item \revC{4U-0.50: 4-ASK uniform with $R_{\tc} = 0.50$};
\item 4U-0.75: 4-ASK uniform with $R_{\tc} = 0.75$;
\item 8PS-0.67: 8-ASK PAS with $R_{\tc} = 0.67$;
\item 8PS-0.83: 8-ASK PAS with $R_{\tc} = 0.83$.
\end{enumerate}
The required \acp{SNR} to operate at the respective \acp{SE} are summarized in Table~\ref{tab:ops_1.5}.

\begin{table}[t]
\centering
\caption{Operating modes and their Shannon limits for \ac{SE} = \SI{1.5}{\bpcu}.}
\label{tab:ops_1.5}
\begin{tabular}{lcc}
\toprule
Mode & $R_\ttx$ [\si{\bpcu}] & $R_{\tbmd}^{-1}(R_\ttx)$ [\si{dB}]\\
\midrule
\revC{4U-0.50} & \revC{\num{1.0}} &\revC{\num{5.2803}}\\
4U-0.75 & \num{1.5}& \num{9.3084}\\
8PS-0.67 &\num{1.5} &\num{8.5334}\\
8PS-0.83 & \num{1.5}&\num{8.5606}\\
\bottomrule
\end{tabular}
\end{table}

As \ac{FEC} codes we consider (asymptotically) regular, protograph-based \ac{SC-LDPC} codes with \ac{VN} degrees $d_{\mttv} = 4$ and $d_{\mttv} = 6$, and design code rates $R_{\tc} \in \{2/3, 3/4, 5/6\}$.  

The submatrices $\vB_i$ in \eqref{eq:sc_B} are given by
\begin{equation}
\vB_i = \underbrace{(1 \quad 1 \quad \ldots \quad 1)}_{d_\tc}, \quad i = 0,\dotsc,\mu,
\end{equation}
where $\mu= d_{\mttv}-1$. The corresponding right-unterminated ensembles are referred to via their basematrices as $\vB^{d_{\mttv},d_{\mttc}}$
in the following.

\subsection{Asymptotic Decoding Thresholds}

The decoding thresholds in \revC{Tables~\ref{tab:cmp_thresholds_sc_4ask_10}}, \ref{tab:cmp_thresholds_sc_4ask_15} and \ref{tab:cmp_thresholds_sc_8ask_15} were obtained for window decoding and a window size of $W = 15$ spatial positions, using the procedure presented in Sec.~\ref{sec:tmp_den_evo_window}. \revA{We use $T = 1.3$ as a threshold parameter for \ac{BMP}, \ac{TMP}, and \ac{QMP} in all numerical evaluations.} A maximum number of \num{1000} iterations per window are performed. These parameters were chosen to depict the absolute performance limits. Increasing the window size did not further affect the numerical results. We conclude that the performance of a block-based decoder is similar. For uniform signaling, we use a consecutive bit mapping of the \ac{BMD} bit channel to each protograph \ac{VN}, i.e., for $2^m$-\ac{ASK} we have 
\FS{\begin{align}
\cV_{\tp}^{(1)} &= \{\mttV_1,\mttV_{1+m},\mttV_{1+2m},\dotsc,\mttV_{n_\tP-(m-1)}\}\\
\vdots& \nonumber\\
\cV_{\tp}^{(m)} &= \{\mttV_{m},\mttV_{2m},\mttV_{3m},\dotsc,\mttV_{n_\tP}\}.
\end{align}
For PAS, we must take into account that bit-level one (representing the sign of the constellation points~\cite{bocherer_bandwidth_2015}) is mainly formed by parity bits and has to be placed accordingly. We choose
\begin{align}
\cV_{\tp}^{(1)} &= \{\mttV_{(n_\tp/m)\cdot(m-1) + 1},\mttV_{(n_\tp/m)\cdot(m-1) + 2},\dotsc,\mttV_{n_\tP}\}\\
\cV_{\tp}^{(2)} &= \{\mttV_1,\mttV_{m},\mttV_{2m-1},\dotsc,\mttV_{(n_\tP/m-1)\cdot(m-1)+1}\}\\
\vdots& \nonumber\\
\cV_{\tp}^{(m)} &= \{\mttV_{m-1},\mttV_{2(m-1)},\mttV_{3(m-1)},\dotsc,\mttV_{(n_\tp/m)\cdot(m-1) }\}.
\end{align}
These mappings are repeated for each spatial position.
}

The decoding threshold for full BP decoding is obtained via discretized \ac{DE}~\cite{chung_design} with $8$-bit quantization and a dynamic range of the soft-information of $[-16,+16]$. Increasing the resolution had no further effect. The \ac{DM} rate for the \ac{PS} modes was chosen according to \eqref{eq:Rdm}, \eqref{eq:Rtx_PAS} and the output symbols have an \ac{MB} distribution~\cite{kschischang_pasupathy_maxwell} with corresponding entropy.

\begin{table}[t]
\centering
\caption{Decoding thresholds in \textnormal{\si{dB}} for 4-ASK uniform and an \ac{SE} of \SI{1.0}{bpcu}.}
\begin{tabular}{lllll}
\toprule
$\vB$ & $\SNR_{\tth}^{\text{full}}$  & $\SNR_{\tth}^{\text{BMP}}$ & $\SNR_{\tth}^{\text{TMP}}$ & $\SNR_{\tth}^{\text{QMP}}$\\
\midrule
$\vB^{4,8}$ & \num{5.36}   &  \num{7.75} & \num{6.50} & \num{6.26}\\
\bottomrule
\end{tabular}
\label{tab:cmp_thresholds_sc_4ask_10}
\end{table}

\begin{table}[t]
\centering
\caption{Decoding thresholds in \textnormal{\si{dB}} for 4-ASK uniform and an \ac{SE} of \SI{1.5}{bpcu}.}
\begin{tabular}{lllll}
\toprule
$\vB$ & $\SNR_{\tth}^{\text{full}}$  & $\SNR_{\tth}^{\text{BMP}}$ & $\SNR_{\tth}^{\text{TMP}}$ & $\SNR_{\tth}^{\text{QMP}}$\\
\midrule
$\vB^{4,16}$ & \num{9.41}   &  \num{10.89}                & \num{10.11} & \num{10.00}\\
$\vB^{6,24}$ & \num{9.34}   &  \num{10.72}                &\num{10.0} &\num{9.88}\\
\bottomrule
\end{tabular}
\label{tab:cmp_thresholds_sc_4ask_15}
\end{table}

\begin{table}[t]
\centering
\caption{Decoding thresholds in \textnormal{\si{dB}} for 8-ASK PS and an \ac{SE} of \SI{1.5}{bpcu}.}
\begin{tabular}{lllll}
\toprule
$\vB$ & $\SNR_{\tth}^{\text{full}}$  & $\SNR_{\tth}^{\text{BMP}}$ & $\SNR_{\tth}^{\text{TMP}}$ & $\SNR_{\tth}^{\text{QMP}}$\\
\midrule
$\vB^{4,12}$ & \num{8.65}   &  \num{10.81}                & \num{9.68} & \num{9.50}\\
$\vB^{4,24}$ & \num{8.67}   &  \num{10.06}                & \num{9.33} & \num{9.23}\\
$\vB^{6,18}$ & \num{8.57}   &  \num{10.62}                &\num{9.55} &\num{9.37}\\
$\vB^{6,36}$ & \num{8.59}   &  \num{9.88}                &\num{9.21} &\num{9.10}\\
\bottomrule
\end{tabular}
\label{tab:cmp_thresholds_sc_8ask_15}
\end{table}

As expected from~\cite{kudekar_threshold_2011}, we see in Tables~\ref{tab:cmp_thresholds_sc_4ask_10}--\ref{tab:cmp_thresholds_sc_8ask_15} that the regular ensembles under full \ac{BP} decoding are able to come close (within a few hundreds of a \si{dB}) to the theoretic limits for the specific signaling modes. 
\revC{In previous works~\cite{lechner_analysis_2012,BenY1902:Protograph}, the authors observed that quantized message passing decoders have a smaller gap to the achievable rate limit for high \ac{FEC} code rates codes. This is also reflected in the following results.

While \ac{BMP}, \ac{TMP}, and \ac{QMP} have gaps of \SI{2.39}{dB}, \SI{1.14}{dB}, and \SI{0.9}{dB} to the unquantized BP threshold for 4U-0.50 (i.e., for $R_\tc = 1/2$), the gaps are only \SI{1.48}{dB}, \SI{0.70}{dB}, and \SI{0.59}{dB} for 4U-0.75 (i.e., for $R_\tc = 3/4$). The gain of \ac{TMP} over \ac{BMP}, i.e., using two bits instead of one, is significant and ranges from \SIrange{0.7}{1.25}{dB} depending on the signaling mode and code ensemble.  The gain of \ac{QMP} over \ac{TMP} is particularly pronounced for low code rates (\SI{0.24}{dB} for 4U-0.50) and decreases for higher code rates to about \SI{0.1}{dB}. We note that these gains
can be obtained at no increase in data flow. This observation has a particular implication for \ac{PAS}, where the same transmission rate can be obtained with different \ac{FEC} code rates by adjusting the signaling distribution: Going from a rate 2/3 to a rate 5/6 code ($\vB^{4,12}$ vs. $\vB^{4,24}$) decreases the decoding threshold by \SI{0.75}{dB} (BMP), \SI{0.35}{dB} (TMP) and \SI{0.27}{dB} (QMP). This is in contrast to full \ac{BP}, where the decoding threshold even slightly deteriorates.
Uniform signaling does not allow this flexibility, as the constellation order and \ac{FEC} code rate directly determine the transmission rate.}

In Table~\ref{tab:cmp_thresholds_surrogate} we show the thresholds for the $\vB^{4,16}$ (uniform) and $\vB^{4,12}$ (PS) ensembles obtained via the surrogate approach of Sec.~\ref{sec:initialization_de_tmp}. We observe that the decoding thresholds numerically coincide.

\begin{table}[t]
\centering
\caption{Decoding thresholds in \textnormal{\si{dB}} via surrogates of selected ensembles of Table~III and Table~IV.}
\begin{tabular}{llll}
\toprule
$\vB$ & $\SNR_{\tth}^{\text{BMP}}$ & $\SNR_{\tth}^{\text{TMP}}$ & $\SNR_{\tth}^{\text{QMP}}$\\
\midrule
$\vB^{4,16}$ &  \num{10.89}   & \num{10.11} & \num{10.0}\\
$\vB^{4,12}$ &  \num{10.81}   & \num{9.68} & \num{9.50}\\
\bottomrule
\end{tabular}
\label{tab:cmp_thresholds_surrogate}
\end{table}

\subsection{Finite Length Simulations}

\begin{figure*}[ht]
\centering
\footnotesize
\subfloat[][Uniform: 4U-0.75]{\includegraphics{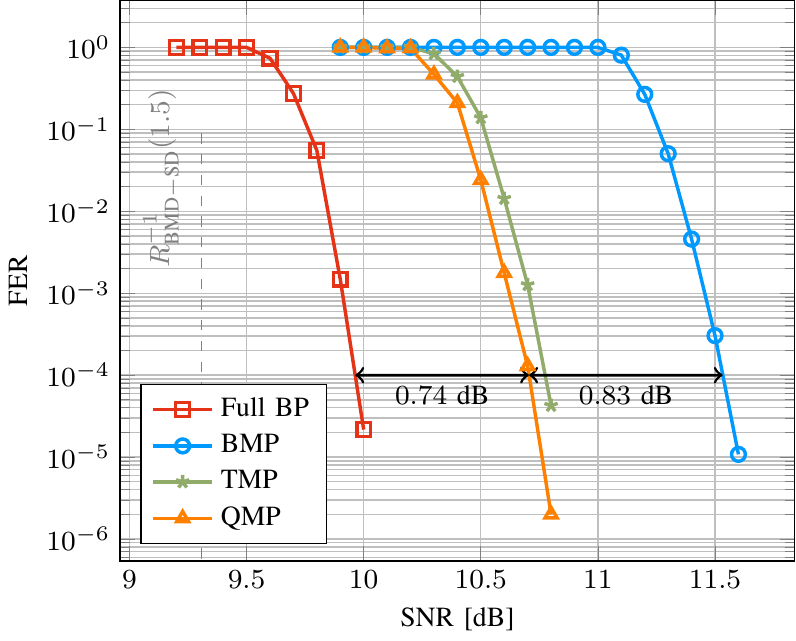}}
\subfloat[][PAS: 8PS-0.83]{\includegraphics{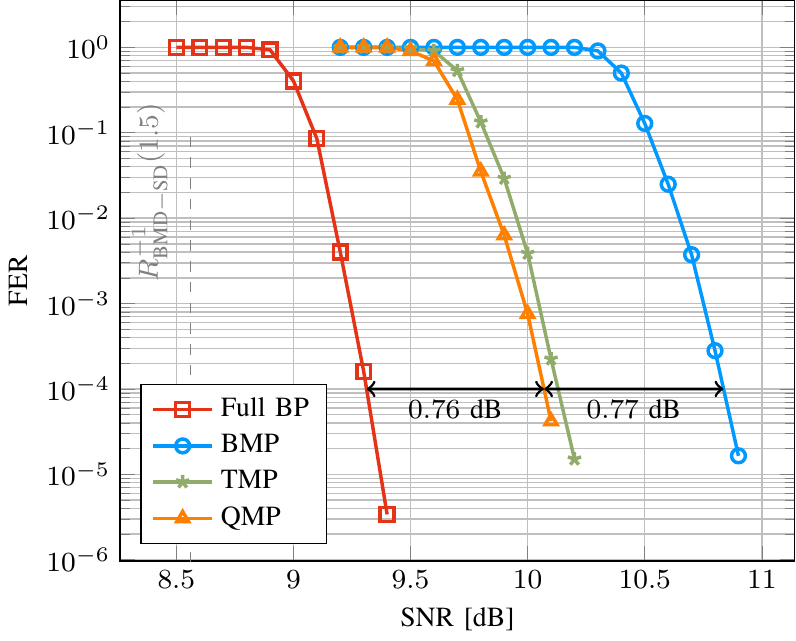}}
\caption{FER simulation results for uniform (a) and PAS (b) signaling and an \ac{SE} of \SI{1.5}{bpcu}.}
\label{fig:coded_results}
\end{figure*}

We validate our asymptotic findings by finite length simulations with a block-based decoder for the 4U-0.75 and 8PS-0.83 signaling modes in Fig.~\ref{fig:coded_results}. We use terminated SC-LDPC codes with $S=50$ spatial positions and an overall blocklength of $n_{\tc} = \SI{60000}{bits}$. The resulting code rates are $0.735$ ($\vB^{4,16}$) and $0.8233$ ($\vB^{4,24}$) according to \eqref{eq:sc_coderate} with lifting factors of $Q=300$ and $Q = 200$, respectively. We used cyclic liftings and girth optimization techniques to ensure a minimum girth of eight. Because of the termination, the effective \ac{SE} is \SI{1.47}{\bpcu}. \revB{The weighting factors were chosen as calculated by the \ac{DE} analysis at the respective decoding threshold.}

For both cases, \ac{QMP} gains about \SI{0.8}{dB} compared to \ac{BMP}. As predicted by \ac{DE}, the performance of \ac{QMP}  improves over \ac{TMP} in the order of about \SI{0.1}{dB}. The gap of QMP to full \ac{BP} decoding is about \SI{0.75}{dB} at a \ac{FER} of \num{e-4}.

\section{Conclusion}
\label{sec:conclusions}

\FS{Reducing the internal \ac{LDPC} decoder data flow is essential for hardware implementations targeting application specific integrated circuits and constitutes an important prerequisite for high throughput decoders.} In this work, we extensively compared one and two bit quantized \ac{BP} decoding algorithms for higher order modulation and \ac{PS}. For this, we also introduced a novel two-bit message passing decoding algorithm with a four-ary message alphabet that improves upon the previous \ac{TMP} approach \revC{by up to \SI{0.24}{dB} for low code rates} while having the same internal decoder dataflow.  We showed how \ac{DE} must be modified for the quantized message passing algorithms to account for higher order modulation and \ac{PS}. The results were applied to protograph-based \ac{SC-LDPC} codes to show the potential of the decoding algorithms for high throughput optical \ac{FEC} solutions. Finite length simulation results reflected the predicted gains by \ac{DE}.

\section*{Acknowledgement}
The authors would like to thank Gerhard Kramer for many helpful discussions.

\appendices
\section{Scaling of CN Messages}
We motivate the scaling of \ac{CN} messages with the scaling factor $ w$. Consider transmission of a binary symbol $x$ over two  channels with transition probabilities $P_{Y_1|X}$ and $P_{Y_2|X}$. Upon observing $y_1$ and $y_2$, in order to perform a maximum likelihood decision we may compute
\begin{align}
l(y_1,y_2) =   \underbrace { \log \frac{P_{Y_1|X}(y_1|0)}{P_{Y_1|X}(y_1|1)}}_{l_1 (y_1)} + \underbrace{\log \frac{P_{Y_2|X}(y_2|0)}{P_{Y_2|X}(y_2|1)}}_{l_2(y_2)} .
\end{align}
We apply the same principle at each \ac{VN} processor. Thus, let the first channel be the communication channel for which $l_1$ is available.  A message sent from a \ac{CN} to a \ac{VN} can be modeled as the observation of the \ac{RV} $X$ after transmission over a binary-input $M$-ary output discrete memoryless extrinsic channel \cite{ashikhmin_EXIT}. For an observation $y_2$,   $l_2$ is computed from the transition probabilities of the extrinsic channel, for which accurate estimates can obtained by \ac{DE} in the large block length regime.

Consider \ac{QMP} as an example. The $4$-ary extrinsic channel output alphabet is $\cY=\{\pm \levH, \pm \levL\}$, and we have
\begin{equation}
l_2(y) = \begin{cases}
\log \frac{P_{Y_2|X}( \levH|0)}{P_{Y_2|X}(\levH|1)}, & y=\levH\\
\log \frac{P_{Y_2|X}(- \levH|0)}{P_{Y_2|X}(- \levH|1)}, & y=-\levH\\
\log \frac{P_{Y_2|X}( \levL|0)}{P_{Y_2|X}(\levL|1)}, & y= \levL\\
\log \frac{P_{Y_2|X}(- \levL|0)}{P_{Y_2|X}(- \levL|1)}, & y=- \levL.
\end{cases}
\end{equation} 
Similarly to \eqref{eq:def_symmetry_channel}, we require that the extrinsic channel fulfills the symmetry constraint
\begin{align}
P_{Y_2|X}( \levH|0)=P_{Y_2|X}(- \levH|1)\\
P_{Y_2|X}( \levL|0)=P_{Y_2|X}(- \levL|1).
\end{align}
Thus, we have
\begin{equation}
|l_2(y)| = \begin{cases}
\log \frac{P_{Y_2|X}( \levH|0)}{P_{Y_2|X}(\levH|1)}, & y=\pm \levH\\
\log \frac{P_{Y_2|X}(\levL|0)}{P_{Y_2|X}( \levL|1)}, & y=\pm \levL.
\end{cases}
\label{eq:L2}
\end{equation} 
Above, we assumed that $P_{Y_2|X}( \levH|0)>P_{Y_2|X}( \levH|1)$ and $P_{Y_2|X}(\levL|0)>P_{Y_2|X}( \levL|1)$. We have
\begin{align}
l(y)=l_1+\sign(y) |l_2(y)|. \label{eq:L}
\end{align}
As in \eqref{eq:L}, the \ac{VN} update in Algorithm~\ref{alg:qmp}  performs a weighting of the extrinsic \ac{CN} messages. 
Observe that the weighting factor $w$ in \eqref{eq:DLej} and \eqref{eq:DHej} is defined as $|l_2(y)|$ in \eqref{eq:L2}.


\end{document}